\documentclass[prd,showpacs]{revtex4}

\usepackage{graphicx}
\usepackage{dcolumn}
\usepackage{bm}

\begin{document}
\renewcommand{\theequation}{\thesection.\arabic{equation}}

\title{Spin-Gravity Coupling and Gravity-Induced Quantum Phases\\}

\author{Giorgio Papini$^{a,b,c}$}
 \altaffiliation[Electronic address:]{papini@uregina.ca}
  \address{$^a$Department of Physics, University of Regina, Regina, SASK, S4S 0A2, Canada}
 \address{$^b$Prairie Particle Physics Institute, Regina, SASK,
S4S 0A2, Canada}
  \address{$^c$International Institute for Advanced Scientific Studies, 89019 Vietri sul Mare (SA), Italy}

\date{\today}

\begin{abstract}
External gravitational fields induce phase factors in the wave
functions of particles. The phases are exact to first order in the
background gravitational field, are manifestly covariant and gauge
invariant and provide a useful tool for the study of spin-gravity
coupling and of the optics of particles in gravitational or inertial
fields. We discuss the role that spin-gravity coupling plays in
particular problems.
\end{abstract}


\maketitle

\section{\label{sec:1}Introduction}

\setcounter{equation}{0}

The study of the interaction of spin with inertia and gravity has
received a strong impulse from the work of Bahram Mashhoon
\cite{mashh1,mashh2,mashh3,mashh4}. His work has stimulated the
research on which we report below.

Covariant wave equations for scalar and vector bosons, for spin-1/2
fermions \cite{cai1,cai2,singh,punzi} and spin-2 \cite{pap3}
particles can be solved exactly to first order in the metric
deviation. The background gravitational and inertial fields appear
in the solutions as phase factors multiplying the wave function of
the corresponding field-free equations. The phases can be calculated
with ease for most metrics.

We summarize the solutions for vector, tensor bosons and fermions in
Sections II, III and IV. In the same sections we also extract the
spin-gravity interaction \cite{hehl} from the gravity-induced
phases. The optics of the particles is derived in Sections V. In
Sections VI-VIII we discuss the relevance of the Mashhoon coupling
to muon $ g-2$ experiments, discrete symmetries and neutrino
helicity transitions. The conclusions are contained in Section IX.

\section{\label{sec:2}Solution of the spin-1 wave equation}
\setcounter{equation}{0}

Photons in gravitational fields are described by the Maxwell-de Rahm
equations \cite{misn}
\begin{equation}\label{a}
  \nabla_\alpha \nabla^\alpha A_\mu - R_{\mu\alpha} A^\alpha = 0\,,
\end{equation}
which reduce to Maxwell equations
\begin{equation}\label{b}
  \nabla_\alpha \nabla_\alpha A_\mu = 0\,
\end{equation}
when $ R_{\mu\alpha}=0 $, or when, as in lensing, the wavelength $
\lambda$ of $ A^{\alpha}$ is much smaller than the typical radius of
curvature of the gravitational background.

In (\ref{a}) and (\ref{b}) $\nabla_\alpha$ indicates covariant
differentiation. We use units $ \hbar=c=1$.

In what follows we consider only (\ref{b}) and its generalization
to massive, charge-less, spin-1 particles
\begin{equation}\label{c}
  \nabla_\alpha \nabla^\alpha A_\mu +m^2 A_\mu = 0\,,
\end{equation}
in the hope that among the atoms and molecules to be used in
interferometry some indeed satisfy (\ref{c}) \cite{borde}.

Following previous work \cite{cai1}-\cite{cai2}, we show below that
equations (\ref{b}) and (\ref{c}) can be solved exactly to first
order in the metric deviation
$\gamma_{\mu\nu}=g_{\mu\nu}-\eta_{\mu\nu}$, where
$|\gamma_{\mu\nu}|\ll 1$ and $\eta_{\mu\nu}$ is the Minkowski metric
of signature -2.

To first order in $\gamma_{\mu\nu}$, (\ref{b}) and (\ref{c}) become
\begin{eqnarray}\label{d}
  \nabla_\nu \nabla^\nu A_\mu \simeq
  (\eta^{\sigma\alpha}-\gamma^{\sigma\alpha})A_{\mu,
  \alpha\sigma}-(\gamma_{\sigma\mu,\nu}+\gamma_{\sigma\nu,\mu}-\gamma_{\mu\nu,\sigma})
  A^{\sigma,\nu}-\frac{1}{2}\gamma_{\sigma\mu,\nu}^{\,\,\,\,\,\,\,\,\nu}A^\sigma &=&
  0\,, \\
    (\nabla_\nu \nabla^\nu+m^2)A_\mu \simeq
  (\eta^{\sigma\alpha}-\gamma^{\sigma\alpha})A_{\mu,\alpha\sigma}-
  (\gamma_{\sigma\mu,\nu}+\gamma_{\sigma\nu,\mu}-\gamma_{\mu\nu,\sigma})
  A^{\sigma,\nu}-\frac{1}{2}\gamma_{\sigma\mu,\nu}^{\,\,\,\,\,\,\,\,\nu}A^\sigma +m^2 A_\mu &=&
  0\,, \label{e}
\end{eqnarray}
where ordinary differentiation of a quantity $\Phi$ is equivalently
indicated by $\Phi_{,\alpha}$ or $\partial_\alpha \Phi$. In deriving
(\ref{d}) and (\ref{e}), we have used the Lanczos-De Donder gauge
condition
\begin{equation}\label{f}
  \gamma_{\alpha\nu,}^{\,\,\,\,\,\,\,
  \nu}-\frac{1}{2}\gamma_{\sigma,\alpha}^\sigma = 0\,.
\end{equation}
In the massless case, the field $A_\mu(x)$ satisfies the condition
\begin{equation}\label{g}
  \nabla_\mu A^\mu=0\,.
\end{equation}
It is convenient to impose (\ref{g}) also in the case of a massive
particle. Equations (\ref{b}) and (\ref{c}) can be handled
simultaneously. Their solution is
\begin{eqnarray}
A_\mu(x) & \simeq & e^{-i\xi}a_\mu(x)\approx (1-i\xi)a_\mu(x) \nonumber\\
 &=&
a_{\mu}(x)-\frac{1}{4}\int_{P}^{x}dz^{\lambda}(\gamma_{\alpha\lambda,\beta}(z)-
\gamma_{\beta\lambda,\alpha}(z))[(x^{\alpha}-z^{\alpha})\partial^{\beta}a_{\mu}(x)
- (x^{\beta}-z^{\beta})\partial^{\alpha}a_{\mu}(x)] \nonumber \\
 & & +\frac{1}{2}\int_{P}^{x}dz^{\lambda} \gamma_{\alpha\lambda}(z)
\partial^{\alpha}a_{\mu}(x)\nonumber \\
 & & -\frac{1}{2}\int_{P}^{x}dz^{\lambda}(\gamma_{\mu\lambda,
 \sigma}(z)-
 \gamma_{\sigma\mu,\lambda}(z)-\gamma_{\sigma\lambda,\mu}(z))
 a^{\sigma}(x)\,, \label{h}
\end{eqnarray}
where $a_\mu$ satisfies the equation $\partial_\nu\partial^\nu
a_\mu =0$ in the case of (\ref{d}), and $(\partial_\nu\partial^\nu
+ m^2)a_\mu =0$ when (\ref{h}) is a solution of (\ref{e}). In
(\ref{h}) $ P$ is a fixed reference point and $ x $ a generic
point along the particle's worldline. We can prove that (\ref{h})
is an exact solution to first order in $\gamma_{\mu\nu}$ by
straightforward differentiation.

The first two integrals in (\ref{h}) represent by themselves a
solution of the Klein-Gordon equation $(\nabla_\mu \nabla^\nu +
m^2)\phi =0$. The additional terms are related to spin. In fact
(\ref{h}) can be re-written in the form
\begin{equation}\label{l}
  A_\mu = a_\mu -\frac{1}{2}\int_P^x dz^\lambda
  (\gamma_{\alpha\lambda,\beta}-\gamma_{\beta\lambda,\alpha})(x^\alpha-z^\alpha)\partial^\beta
  a_\mu +\frac{1}{2}\int_P^x dz^\lambda
  \gamma_{\alpha\lambda}\partial^\alpha a_\mu-\frac{i}{2}\int_P^x
  dz^\lambda
  (\gamma_{\alpha\lambda,\beta}-\gamma_{\beta\lambda,\alpha})S^{\alpha\beta}a_\mu
   \end{equation}
 \[
  +\frac{i}{2}\int_P^x dz^\lambda
  \gamma_{\alpha\beta,\lambda}T^{\alpha\beta}a_\mu (x)\,,
 \]
where
\begin{equation}\label{m}
  (S^{\alpha\beta})_{\mu\nu}=-\frac{i}{2}(\delta^\alpha_\mu
  \delta_\nu^\beta -\delta_\mu^\beta \delta_\nu^\alpha)\,,
  \quad
 (T^{\alpha\beta})_{\mu\nu}=-\frac{i}{2}(\delta^\alpha_\mu
  \delta_\nu^\beta +\delta_\mu^\beta \delta_\nu^\alpha)\,.
\end{equation}
The rotation matrices $S_i=2i\epsilon_{ijk}S^{jk}$ satisfy the
commutation relation $[S_i, S_j]=i\epsilon_{ijk}S_k$.

By applying Stokes theorem to the r.h.s. of (\ref{h}) we find
\begin{equation}\label{n}
  A_\mu = \left(1-\frac{i}{4}\oint d\tau^{\sigma\delta}R_{\sigma\delta \alpha
  \beta}J^{\alpha\beta}\right) a_\mu \,,
\end{equation}
where $J^{\alpha\beta}=L^{\alpha\beta}+S^{\alpha\beta}$ is the total
angular momentum of the spin-1 particle, $R_{\mu\nu\alpha\beta}=1/2
(\gamma_{\mu\beta,\nu\alpha}+\gamma_{\nu\alpha,\mu\beta}-\gamma_{\mu\alpha,\nu\beta}-
\gamma_{\nu\beta,\mu\alpha})$ is the linearized Riemann tensor, and
$\tau$ is the surface bound by the closed path along which the
integration is performed.

The weak field approximation $ g_{\mu\nu}=\eta_{\mu\nu}+
\gamma_{\mu\nu}$ does not fix the reference frame completely. The
transformations of coordinates $ x_{\mu}\rightarrow
x_{\mu}+\xi_{\mu}$, with $ \xi_{\mu}(x) $ also small of first
order, are still allowed and lead to the "gauge" transformations $
\gamma_{\mu\nu}\rightarrow
\gamma_{\mu\nu}-\xi_{\mu,\nu}-\xi_{\nu,\mu}$. Equation (\ref{n})
therefore indicates that solution (\ref{h}) is covariant and also
gauge invariant. It also follows from (\ref{n}) that the term
containing $T^{\alpha\beta}$ in (\ref{l}) does not contribute to
integrations over closed paths, behaves as a gauge term and may
therefore be dropped.

The spin-gravity coupling is contained in the third term on the
r.h.s. of (\ref{l}). Its time integral part is $
\xi_{sr}=-\frac{1}{4} \int^{x}_{P} dz^{0} (\gamma_{\alpha 0,
\beta}(z) - \gamma_{\beta 0, \alpha}(z))S^{\alpha \beta} $. Since
for rotation $ \gamma_{0i}  = (-\Omega y, \Omega x, 0)  $, one gets
$  \xi_{sr}= -\frac{1}{2} \int dz^{0} \gamma_{i 0, j}(z) S^{ij} =
\int dt \Omega S_{z} $, where $S_{z} = S^{12}$. In general, one may
write $ \int dt \bm{\Omega} \cdot \bm{S}$, which must now be applied
to a solution of the field free equations. One finds $
E^{\prime}_{\pm} = E + \bm{\Omega} \cdot \bm{S} $ and, for particles
polarized parallel or antiparallel to $\bm{\Omega}$, $
E^{\prime}_{\pm} = E \pm \hbar \Omega $, as in
\cite{mashh}.$E^{\prime}$ is the energy observed by the co-rotating
observer.

\section{Solution of the spin-2 wave equation}
\setcounter{equation}{0}

For spin-$2$ fields, the simplest equation of propagation is derived
in \cite{misn} and is given by
\begin{equation}\label{MTW}
\nabla_{\alpha}\nabla^{\alpha}\Phi_{\mu\nu}+ 2
R_{\alpha\mu\beta\nu}\Phi^{\alpha\beta}=0\,.
\end{equation}
In lensing, the second term in (\ref{MTW}) may be neglected when the
wavelength $ \lambda$ associated with $ \Phi_{\mu\nu}$ is smaller
than the typical radius of curvature of the gravitational background
\cite{tak}.

We consider here massless or massive spin-$2$ particles described by
the equation \cite{pap3}
\begin{equation}\label{spin2}
\nabla_{\alpha}\nabla^{\alpha}\Phi_{\mu\nu}+ m^2 \Phi_{\mu\nu}= 0.
\end{equation}

To first order
in $ \gamma_{\mu\nu}$, (\ref{spin2}) can be written in the form
\begin{equation} \label{wf}
\left(\eta^{\alpha\beta}-\gamma^{\alpha\beta}\right)\partial_{\alpha}\partial_{\beta}\Phi_{\mu\nu}+
R_{\sigma\mu}\Phi_{\nu}^{\sigma}+
R_{\sigma\nu}\Phi_{\mu}^{\sigma}-2\Gamma_{\mu\alpha}^{\sigma}\partial^{\alpha}\Phi_{\nu\sigma}-
2\Gamma_{\nu\alpha}^{\sigma}
\partial^{\alpha}\Phi_{\mu\sigma}+ m^2 \Phi_{\mu\nu} = 0,
\end{equation}
where
$R_{\mu\beta}=-(1/2)\partial_{\alpha}\partial^{\alpha}\gamma_{\mu\beta}$
is the linearized Ricci tensor of the background metric and $
\Gamma_{\sigma\mu,\alpha}=1/2\left(\gamma_{\alpha\sigma,\mu}+\gamma_{\alpha\mu,\sigma}-
\gamma_{\sigma\mu,\alpha}\right)$ is the corresponding Christoffel
symbol of the first kind.

It is easy to prove, by direct substitution, that a solution of
(\ref{wf}), exact to first order in $ \gamma_{\mu\nu}$, is
represented by
\begin{eqnarray}\label{solution}
\Phi_{\mu\nu}=
\phi_{\mu\nu}-\frac{1}{4}\int_{P}^{x}dz^{\lambda}\left(\gamma_{\alpha\lambda,\beta}\left(z\right)-
\gamma_{\beta\lambda,\alpha}\left(z\right)\right)\left[\left(x^{\alpha}-z^{\alpha}\right)\partial^{\beta}\phi_{\mu\nu}
\left(x\right)-\left(x^{\beta}-z^{\beta}\right)\partial^{\alpha}\phi_{\mu\nu}\left(x\right)\right]
\\ \nonumber + \frac{1}{2}
\int_{P}^{x}dz^{\lambda}\gamma_{\alpha\lambda}\left(z\right)\partial^{\alpha}\phi_{\mu\nu}\left(x\right)+
\int_{P}^{x}dz^{\lambda}\Gamma_{\mu\lambda,\sigma}\left(z\right)\phi_{\nu}^{\sigma}\left(x\right)+
\int_{P}^{x}dz^{\lambda}\Gamma_{\nu\lambda,\sigma}\left(z\right)
\phi_{\mu}^{\sigma}\left(x\right),
\end{eqnarray}
where $\phi_{\mu\nu}$ satisfies the field-free equation
\begin{equation} \label{ff}
\left(\partial_{\alpha}\partial^{\alpha}+
m^2\right)\phi_{\mu\nu}\left(x\right)=0 \,,
\end{equation}
and the gauge condition (\ref{f}) has been used.

Equation (\ref{solution}) can be written in the form
\begin{eqnarray}\label{sol'}
\Phi_{\mu\nu}\left(x\right)=\phi_{\mu\nu}\left(x\right)+\frac{1}{2}\int_{P}^{x}dz^{\lambda}\gamma_{\alpha\lambda}
\left(z\right)
\partial^{\alpha}\phi_{\mu\nu}\left(x\right)-
\frac{1}{2}\int_{P}^{x}dz^{\lambda}\left(\gamma_{\alpha\lambda,\beta}\left(z\right)-\gamma_{\beta\lambda,\alpha}
\left(z\right)\right)\left[\left(x^{\alpha}-z^{\alpha}\right)\partial^{\beta}\right]\phi_{\mu\nu}+ \\
\nonumber -
\frac{i}{2}\int_{P}^{x}dz^{\lambda}\left(\gamma_{\alpha\lambda,\beta}\left(z\right)-\gamma_{\beta\lambda,\alpha}\right)S^{\alpha\beta}\phi_{\mu\nu}
\left(x\right) - \frac{i}{2}
\int_{P}^{x}dz^{\lambda}\gamma_{\beta\sigma,\lambda}\left(z\right)T^{\beta\sigma}\phi_{\mu\nu}\left(x\right)\,,
\end{eqnarray}
where \begin{eqnarray}\label{ST}
S^{\alpha\beta}\phi_{\mu\nu}&\equiv& \frac{i}{2}
\left(\delta_{\sigma}^{\alpha}\delta_{\mu}^{\beta}\delta_{\nu}^{\tau}-\delta_{\sigma}^{\beta}
\delta_{\mu}^{\alpha}\delta_{\nu}^{\tau}+\delta_{\sigma}^{\alpha}\delta_{\nu}^{\beta}\delta_{\mu}^{\tau}-
\delta_{\sigma}^{\beta}\delta_{\nu}^{\alpha}\delta_{\mu}^{\tau}\right)\phi_{\tau}^{\sigma}
\\ \nonumber
T^{\beta\sigma}\phi_{\mu\nu}&\equiv&
i\left(\delta_{\mu}^{\beta}\delta_{\nu}^{\tau}+\delta_{\nu}^{\beta}\delta_{\mu}^{\tau}\right)\phi^{\sigma}_{\tau}\,.
\end{eqnarray}
From $ S^{\alpha\beta}$ one constructs the rotation matrices $
S_{i}=-2i \epsilon_{ijk}S^{jk}$ that satisfy the commutation
relations $ [S_{i},S_{j}]=i \epsilon_{ijk}S_{k}$. The spin-gravity
interaction is therefore contained in the term
\begin{equation}\label{SG}
\Phi_{\mu\nu}'\equiv
-\frac{i}{2}\int_{P}^{x}dz^{\lambda}\left(\gamma_{\alpha\lambda,\beta}-\gamma_{\beta\lambda,\alpha}\right)
S^{\alpha\beta}\phi_{\mu\nu}\left(x\right)=\frac{1}{2}\int_{P}^{x}dz^{\lambda}\left[\left(\gamma_{\sigma\lambda,\mu}-
\gamma_{\mu\lambda,\sigma}\right)\phi_{\nu}^{\sigma}+\left(\gamma_{\sigma\lambda,\nu}-\gamma_{\nu\lambda,\sigma}\right)
\phi_{\mu}^{\sigma}\right].
\end{equation}
The solution (\ref{solution}) is invariant under the gauge
transformations $ \gamma_{\mu\nu}\rightarrow
\gamma_{\mu\nu}-\xi_{\mu,\nu}-\xi_{\nu,\mu}$. If, in fact, we
choose a closed integration path $ \Gamma $, Stokes theorem
transforms the first three integrals of (\ref{sol'}) into the
gauge invariant result
\begin{equation}\label{phasephi}
\Phi_{\mu\nu}=\left(1-\frac{i}{4}\int_{\Sigma}d\sigma^{\lambda\kappa}R_{\lambda\kappa\alpha\beta}J^{\alpha\beta}\right)
\phi_{\mu\nu}\,,
\end{equation}
where $ \Sigma$ is the surface bound
by $ \Gamma$, and $
J^{\alpha\beta}=L^{\alpha\beta}+S^{\alpha\beta}$ is the total
angular momentum of the particle. For the same path $ \Gamma $ the
integral involving $ T^{\beta\sigma}$ in (\ref{sol'}) vanishes. It
behaves like a gauge term and is therefore dropped.

The helicity-rotation coupling for massless, or massive spin-$2$
particles follows immediately from the $S^{\alpha\beta}$ term in
(\ref{sol'}). In fact, the particle energy is changed by virtue of
its spin by an amount given by the time integral of this spin term
\begin{equation}\label{hrc}
\xi^{hr}=-\frac{1}{2}\int_{P}^{x}dz^{0}\left(\gamma_{\alpha0,\beta}-\gamma_{\beta0,\alpha}\right)S^{\alpha\beta}\,,
\end{equation}
that must then be applied to a solution of (\ref{ff}). For rotation
about the $ x^{3}$-axis, $ \gamma_{0i}=\Omega(y,-x,0) $, we find $
\xi^{hr}=-\int_P^x dz^0 2 \Omega S^3 $ and the energy of the
particle therefore changes by $\pm 2\Omega$, where the factor $\pm
2$ refers to the particle's helicity, as discussed by Ramos and
Mashhoon \cite{ramos}. Equation (\ref{hrc}) extends their result to
any weak gravitational, or inertial field.

The effect of (\ref{SG}) on $ \phi_{\mu\nu}$ can be easily seen in
the case of a gravitational wave propagating in the $x$-direction
and represented by the components $
\phi_{22}=-\phi_{33}=\varepsilon_{22}exp\left[ik\left(t-x\right)\right]$
and $ \phi_{23}=\varepsilon_{23}exp\left[ik\left(t-x\right)\right]$.
For an observer rotating about the $ x$-axis the metric is $
\gamma_{00}=-\Omega^2 r^2\,,
\gamma_{11}=\gamma_{22}=\gamma_{33}=-1\,,\gamma_{0i}=\Omega(0,z,-y)
$. Then the two independent polarizations $ \phi_{23}$ and $
\phi_{22}-\phi_{33} $ are transformed by $ S_{\alpha\beta}$ into $
\Phi_{23}=-2\,\Omega\,
\left(x^{0}-x_P^0\right)\,\left(\phi_{22}-\phi_{33}\right)/2$ and $
1/2\,\left(\Phi_{22}-\Phi_{33}\right)=2\,\Omega(x^0-x_P^0)\,\phi_{23}$.

For closed integration paths and vanishing spin, (\ref{sol'})
coincides with the solution of a scalar particle in a gravitational
field, as expected. This proves the frequently quoted statement
\cite{thorne} that gravitational radiation propagating in a
gravitational background is affected by gravitation in the same way
that electromagnetic radiation is (when the photon spin is
neglected).

\section{The Covariant Dirac equation}
\setcounter{equation}{0}

The behavior of spin-1/2 particles in the presence of a
gravitational field $g_{\mu\nu}$ is determined by the covariant
Dirac equation
\begin{equation}\label{DiracEquation}
  [i\gamma^\mu(x){\cal D}_\mu-m]\Psi(x)=0\,,
\end{equation}
where ${\cal D}_\mu=\nabla_\mu+i\Gamma_\mu (x)$, $\Gamma_{\mu}(x)$
is the spin connection and the matrices $\gamma^{\mu}(x)$ satisfy
the relations $\{\gamma^\mu(x), \gamma^\nu(x)\}=2g^{\mu\nu}$. Both
$\Gamma_\mu(x)$ and $\gamma^\mu(x)$ can be obtained from the usual
constant Dirac matrices by using the vierbein fields $e_{\hat
\alpha}^\mu$ and the relations
\begin{equation}\label{II.2}
  \gamma^\mu(x)=e^\mu_{\hat \alpha}(x) \gamma^{\hat
  \alpha}\,,\qquad
  \Gamma_\mu(x)=-\frac{1}{4} \sigma^{{\hat \alpha}{\hat \beta}}
  e^\nu_{\hat \alpha}e_{\nu\hat{\beta};\, \mu}\,,
\end{equation}
where $\sigma^{{\hat \alpha}{\hat \beta}}=\frac{i}{2}[\gamma^{\hat
\alpha}, \gamma^{\hat \beta}]$.

Equation (\ref{DiracEquation}) can be solved exactly to first order
in
 $\gamma_{\mu\nu}(x)$. This is
achieved by first transforming (\ref{DiracEquation}) into the
equation \cite{cai2},\cite{pap1},\cite{punzi}
\begin{equation}\label{DiracEqTrasf}
  [i{\tilde \gamma}^\nu (x) \nabla_{\nu}-m]{\tilde \Psi}(x)=0\,,
\end{equation}
where
\begin{equation}\label{PsiTilde}
{\tilde \Psi}(x)=S^{-1}\Psi(x)\,,\qquad S(x)=e^{-i\Phi_s(x)}\,,
\qquad \Phi_s(x)={\cal P}\int_P^x dz^\lambda \Gamma_\lambda (z)\,,
\qquad {\tilde \gamma}^{\mu}(x)=S^{-1}\gamma^{\mu}(x) S\,.
\end{equation}
By multiplying (\ref{DiracEqTrasf}) on the left by $(-i{\tilde
\gamma}^\nu (x)\nabla_{\nu}-m)$, we obtain the equation
\begin{equation}\label{KGequation}
  (g^{\mu\nu}\nabla_\mu\nabla_\nu+m^2){\tilde \Psi}(x)=0\,,
\end{equation}
whose solution
\begin{equation}\label{ExactSolution}
  {\tilde \Psi}(x)=e^{-i\Phi_G(x)}\Psi_0(x)\,,
\end{equation}
is exact to first order. The operator $\hat{\Phi}_G(x)$ is defined
as
\begin{equation}\label{PhiG}
  \hat{\Phi}_{G}=-\frac{1}{4}\int_P^xdz^\lambda\left[\gamma_{\alpha\lambda,
  \beta}(z)-\gamma_{\beta\lambda, \alpha}(z)\right]\hat{L}^{\alpha\beta}(z)+
  \frac{1}{2}\int_P^x dz^\lambda\gamma_{\alpha\lambda}\hat{k}^\alpha\,,
\end{equation}
\[
 [\hat{L}^{\alpha\beta}(z), \Psi_0(x)]=\left((x^\alpha-z^\alpha)\hat{k}^\beta-
 (x^\beta-z^\beta)\hat{k}^\alpha\right)\Psi_0(x)\,, \qquad
 [\hat{k}^\alpha, \Psi_0(x)]=i\partial^\alpha\Psi_0\,,
 \]
and $\Psi_0(x)$ satisfies the usual flat space-time Dirac
equation. $\hat{L}_{\alpha\beta}$ and $\hat{k}^\alpha$ are the
angular and linear momentum operators of the particle. It follows
from (\ref{ExactSolution}) and (\ref{PsiTilde}) that the solution
of (\ref{DiracEquation}) can be written in the form
\begin{equation}\label{PsiSolution}
  \Psi(x)=e^{-i\Phi_s}\left(-i{\tilde \gamma}^\mu(x)\nabla_\mu
  -m\right)e^{-i\Phi_G}\, \Psi_0(x)\,,
\end{equation}
and also as
\begin{equation}\label{PsiSolution2}
  \Psi(x)=-\frac{1}{2m}\left(-i\gamma^\mu(x){\cal
  D}_\mu-m\right)e^{-i\Phi_T}\Psi_0(x)\,,
\end{equation}
where $\Phi_T=\Phi_s+\Phi_G$ is of first order in
$\gamma_{\alpha\beta}(x)$. The factor $ -1/2m $ on the r.h.s. of
(\ref{PsiSolution2}) is required by the condition that both sides of
the equation agree when the gravitational field vanishes.

It is useful to re-derive some known results from the covariant
Dirac equation. On multiplying (\ref{DiracEquation}) on the left by
$(-i\gamma^\nu(x){\cal D}_\nu-m)$ and using the relations
\begin{equation}\label{relation1}
  \nabla_\mu\Gamma_\nu(x)-\nabla_\nu\Gamma_\mu(x)+i[\Gamma_\mu(x),
  \Gamma_\nu(x)]=-\frac{1}{4}\sigma^{\alpha\beta}(x)R_{\alpha\beta\mu\nu}\,,
\end{equation}
and
\begin{equation}\label{relation2}
  [{\cal D}_\mu, {\cal D}_\nu]=-\frac{i}{4}\,
  \sigma^{\alpha\beta}(x)R_{\alpha\beta\mu\nu}\,,
\end{equation}
we obtain the equation
\begin{equation}\label{KGEq+R}
  \left(g^{\mu\nu}{\cal D}_\mu{\cal
  D}_\nu-\frac{R}{4}+m^2\right)\Psi(x)=0\,.
\end{equation}
In (\ref{relation2}) and (\ref{KGEq+R})
$\sigma^{\alpha\beta}(x)=(i/2)[\gamma^\alpha(x), \gamma^\beta(x)]$
and $R$ is the Ricci scalar.

On applying Stokes theorem to a closed space-time path $C$ and
using (\ref{relation1}), we find that $\Phi_T$ changes by
\begin{equation}\label{phase}
  \Delta \Phi_T=-\frac{i}{4}\oint d\tau^{\mu\nu}J^{\alpha\beta}
  R_{\mu\nu\alpha\beta}\,,
\end{equation}
where $J^{\alpha\beta}$ is the total momentum of the particle.
Equation (\ref{phase}) shows that (\ref{PsiSolution2}) is gauge
invariant.

The spin-rotation coupling derived by Mashhoon by extending the
hypothesis of locality can be now derived rigorously from the
solution found.

Choose a cylindrical coordinate ($t$, $r$, $\theta$, $z$) for an
inertial frame $F_{0}$. An observer at rest in a frame
$F^{\prime}$ rotating with a constant angular velocity $\Omega$
relative to $F_{0}$ will follow the world line $ \left(
r=\mbox{const.}, \theta =\mbox{const.} + \Omega t,
z=\mbox{const.}\right)$.  Consider an orthogonal tetrad consisting
of the observer's four-velocity $\lambda^{\mu}_{(0)} =
dx^{\mu}/ds$ and the triad $\lambda^{\mu}_{(i)}$ ($i = 1,2,3$)
normal to the world line. By using the local tetrad \cite{cai2} $
\lambda^{\mu}_{(0)} = ( \gamma, 0, \frac{\gamma \Omega}{c},
0)\,\,, \lambda^{\mu}_{(1)} = ( 0, 1, 0, 0) \,\,,
\lambda^{\mu}_{(2)} = ( \frac{\gamma \Omega r}{c}, 0,
\frac{\gamma}{r}, 0)\,\,, \lambda^{\mu}_{(3)} = ( 0, 0, 0,1) $,
where $\gamma \equiv (1- r^{2} \Omega^{2}/c^{2})^{-1/2}$, one can
construct a vierbein field $h^{\mu}\, _{a}(x)$ along the world
line of the observer $ h^{\mu}\, _{(0)} = (\gamma, 0, \frac{\gamma
\Omega}{c}, 0) \,, h^{\mu}\, _{(1)} = (-\frac{\gamma \Omega r}{c}
\sin \gamma \Omega t, \cos \gamma \Omega t\,, \frac{\gamma}{r}
\sin \gamma \Omega t, 0)\,, h^{\mu}\, _{(2)} = (\frac{\gamma
\Omega r}{c} \cos \gamma \Omega t, \sin \gamma \Omega t,
\frac{\gamma}{r} \cos \gamma \Omega t, 0) \,, h^{\mu}\, _{(3)} =
(0, 0, 0, 1)$. It is then easy to calculate the spinor connection
$\Gamma_{\mu}$. In calculating the energy, only the component
$\Gamma_{0}$ is necessary. By using the Dirac representation for
the $\gamma$-matrices, one obtains $\Gamma_{0} = \frac{\gamma
\Omega}{2c} \sigma_{z}$ and from $ \Phi_{s}$ also $ \exp(-i \int
\Gamma_{0} dz^{0}) \Psi_{0} = \exp(-\frac{1}{2} \int \gamma \Omega
\sigma_{z} dt) \Psi_{0} $, where $\Psi_{0}$ has the usual plane
wave form. Besides the contribution due to the coupling of the
orbital angular momentum to rotation, which gives the Sagnac
effect \cite{cai1}, one obtains the spin-rotation coupling $
E^{\prime} = E + \frac{\hbar}{2} \Omega \sigma_{z}$, and also, for
spin polarizations parallel or antiparallel to the direction of
rotation, one obtains, $ E^{\prime}_{\pm} = E \pm \frac{\hbar}{2}
\Omega $, as shown by Mashhoon. The present result is exact and
follows from the general form of the solution
(\ref{PsiSolution2}). It also agrees with those of Hehl and Ni
\cite{hehlni} and \cite{singh}.

The Mashhoon effect is obviously a prime candidate for experiments
with accelerators and will be discussed at length below.

According to (\ref{phase}), both angular momentum and spin couple to
a weak gravitational field in the same way. This confirms that,
unlike the electromagnetic case, the gyro-gravitational ratio of a
spin-1/2 particle is 1, as shown in
\cite{oliveira,audretsch,kannenberg}. A classical charge $ e$ moving
in a circle with angular momentum $ \vec{L}$ forms a current loop of
magnetic moment $ \vec{M}= -\frac{e \vec{L}}{2 m c}$, which gives
the gyromagnetic factor $ g=1$. The magnetic moment of a charged
particle depends therefore on the ratio $ e/m$ and, for a rotating
object, on the space distributions of charge and mass. For a quantum
particle, the Dirac equation indicates that $ g=2$. The corrections
to $ g=2$ come from quantum electrodynamics where the electron can
be pictured at any instant as a bare particle in interaction with a
cloud of virtual photons. Qualitatively, if the charge remains
associated with the electron, part of the mass energy is carried by
the photon cloud resulting in a slight increase for the value $ e/m
$ of the electron itself.

In the gravitational case, however, the gyro-gravitational ratio of
the spin-1/2 particle is $ g=1 $. This suggests, according to
\cite{oliveira}, that the internal distributions of the
gravitational mass, associated with the interaction, and of the
inertial mass, associated with the angular momentum, equal each
other.

\section{\label{sec:4}Optics}
\setcounter{equation}{0}

\subsection{\label{sec:4A}Lensing}

In the geometrical optics approximation, valid whenever
$|\partial_i\gamma_{\mu\nu}|\ll|k\gamma_{\mu\nu} |$, where $k$ is
the momentum of the particle, the interaction between the angular
momentum of the source and the particle's spin vanishes. This
interaction is quantum mechanical in origin. Then the geometrical
phase $\Phi_G$ is sufficient to reproduce the classical angle of
deflection \cite{lamb}, as it should, because (\ref{PhiG}) coincides
with the first two integrals in (\ref{h}) and (\ref{sol'}). We can
therefore treat photons, gravitons and fermions simultaneously when
spin is neglected.

More detailed calculations involving neutrinos are given in
\cite{punzi}.

If, e.g., we choose a gravitational background represented by the
Lense-Thirring metric \cite{lense}, $\gamma_{00}=2\phi$,
$\gamma_{ij}=2\phi\delta_{ij}$, $\phi=-GM/r$, and $\gamma_{0i}\equiv
h_i = 2GJ_{ij}x^j/r^3$, with $x^j=(x, y, z)$, $r=\sqrt{x^2+y^2+z^2}$
and $J_{ij}$ is related to the angular momentum of the gravitational
source. In particular, if the source rotates with angular velocity
${\bbox \omega}=(0,0,\omega)$, then $h_1=4GMR^2\omega y/5r^3$,
$h_2=-4GMR^2\omega x/5r^3$.

Without loss of generality, we assume that the particles are
massless and propagate along the $z$-direction, hence
$k^\alpha\simeq (k, 0, 0, k)$. Using plane waves for the field free
solution, the phase of the wave equation becomes
\begin{equation}\label{gph}
 \chi =k_{\alpha}x^{\alpha} -\frac{1}{4}\int_P^x dz^\lambda
 (\gamma_{\alpha\lambda,\beta}(z)-\gamma_{\beta\lambda,
 \alpha}(z))[(x^\alpha-z^\alpha)k^\beta-(x^\beta-z^\beta)k^\alpha]+
 \frac{1}{2}\int_P^x dz^\lambda \gamma_{\alpha\lambda}(z)
 k^\alpha\,.
 \end{equation}
We can define the particle momentum as
\begin{equation}\label{photonmomentum}
 \tilde{k}_\alpha = \frac{\partial \chi}{\partial x^\alpha}\,.
\end{equation}
It is easy to show that $ \chi $ satisfies the eikonal equation $
g^{\alpha\beta} \chi_{,\alpha} \chi_{,\beta}=0 $.

 For the Lense-Thirring metric, $\chi$ is given by
\begin{eqnarray}\label{chiLT}
 \chi & \simeq & -\frac{k}{2}\int_P^Q \left[
   (x-x')\phi_{, z'}dx'+(y-y')\phi_{,\, z'}dy'-2[(x-x')\phi_{,\, x'}+(y-y')\phi_{,
   \, y'}]dz'\right]+k\int_P^Q dz' \phi \\
   & & -\frac{k}{2}\int_P^Q \left[\left(
    (x-x')(h_{1, \,z'}-2h_{3,\, x'})+(y-y')(h_{2, \,z'}-2h_{3,\,
    y'})\right)dz' \right. \nonumber \\
    & & \left. \quad -\left((x-x')h_{1,\, x'}+(y-y')h_{1,\, y'}\right)dx'
   -\left((x-x')h_{2,\, x'}+(y-y')h_{2,\, y'}\right)dy'\right] \nonumber \\
  & & +\frac{k}{2}\int_P^Q\left[2h_3dz'+h_1dx'+h_2dy'\right]\,,
\end{eqnarray}
where $P$ is the point at which the particles are generated, and $Q$
is a generic point along their space-time trajectory. The components
of the momentum are therefore
\begin{eqnarray}\label{k1}
 \tilde{k}_1 &=& 2k\int_P^Q  \left(-\frac{1}{2}\frac{\partial \phi}{\partial z}\,dx -
 \frac{1}{2}\frac{\partial h_2}{\partial x}\,dy +\frac{\partial (\phi+h_3)}{\partial x}dz\right)
 -\frac{k}{2}(h_1(Q)-h_1(P))\,,\label{k2}\\
 \tilde{k}_2 &=& 2k\int_P^Q  \left(-\frac{1}{2}\frac{\partial \phi}{\partial z}\,dy +\frac{1}{2}\frac{\partial h_1}{\partial y}\,dx +
 \frac{\partial (\phi+h_3)}{\partial y}dz\right) +\frac{k}{2}(h_2(Q)-h_2(P))\,, \\
  \tilde{k}_3 &=& k(1+\phi+h_3)\,. \label{k3}
\end{eqnarray}
We then have
\begin{equation}\label{kvector}
  {\bf \tilde{k}}={\bf \tilde{k}}_\perp +\tilde{k}_3\, {\bf e}_3\,, \quad {\bf
  \tilde{k}}_\perp=\tilde{k}_1\, {\bf e}_1+\tilde{k}_1\, {\bf e}_2\,,
\end{equation}
where ${\bf \tilde{k}}_\perp$ is the component of the momentum
orthogonal to the direction of propagation of the particles.

Since only phase differences are physical, it is convenient to
choose the space-time path by placing the particle source at
distances that are very large relative to the dimensions of the
lens, and the generic point is located along the $z$ direction. We
therefore replace $Q$ with $z$, where $z\gg x, y$. Using the
expression for $h_{1,2}$ we find that their contribution is
negligible and (\ref{k1})-(\ref{k3}) simplify to
\begin{eqnarray}\label{k1bis}
 \tilde{k}_1 &=& 2k\int_{-\infty}^z  \frac{\partial (\phi+h_3)}{\partial x}dz\,,\\
 \tilde{k}_2 &=& 2k\int_{-\infty}^z  \frac{\partial (\phi+h_3)}{\partial y}dz\,, \label{k2bis} \\
  \tilde{k}_3 &=& k(1+\phi+h_3)\,. \label{k3bis}
\end{eqnarray}
From (\ref{k1bis})-(\ref{k3bis}) we can determine the deflection
angle $\theta$. Let us analyze the case of non-rotating lenses, i.e.
$h_3=0$. We get
\begin{eqnarray}\label{k1tris}
 \tilde{k}_1 &\sim & k\, \frac{2GM}{R^2}\, x\left(1+\frac{z}{r}\right)\,, \\
 \tilde{k}_2 &\sim &  k\, \frac{2GM}{R^2}\, y\left(1+\frac{z}{r}\right) \,,\label{k2tris} \\
  \tilde{k}_3 &=& k(1+\phi+h_3)\,, \label{k3tris}
\end{eqnarray}
where $R=\sqrt{x^2+y^2}$. By defining the deflection angle as
\begin{equation}\label{theta}
  \tan \theta = \frac{\tilde{k}_\perp}{\tilde{k}_3}\,,
\end{equation}
it follows that
\begin{equation}\label{thetak}
  \tan \theta \sim \theta \sim
  \frac{2GM}{R}\left(1+\frac{z}{r}\right)\,.
\end{equation}
In the limit $z\to \infty$ we obtain the usual Einstein result
\begin{equation}\label{thetaM}
  \theta_M\sim \frac{4GM}{R}\,.
\end{equation}

A general expression for the index of refraction $ n $ can also be
derived from (\ref{gph}),(\ref{photonmomentum}) and  $ n =
\tilde{k}/\tilde{k}_0 $.

\subsection{\label{sec:4B}Wave effects in gravitational lensing}

We now consider the propagation of light and gravitational waves in
a background metric represented by $\gamma_{00}=2U(r)$,
$\gamma_{ij}=2 U(r) \delta_{ij}$, where $U(r)=-GM/r$ is the
gravitational potential of the lens and $r$ the distance from $ M $
to the particle. Wave optics effects can be seen by using the type
of double slit arrangement indicated in Fig.1. We will also use a
solution $\partial^\gamma
\partial_\gamma a_\alpha =0$ in the form of a plane wave
$a_\mu = a_\mu^0 e^{-ik_\sigma x^\sigma}$ and neglect spin effects.
This limits the calculation \cite{pap3} of the phase difference to
the first two terms in (\ref{9}) and (\ref{10}). We also assume for
simplicity that $k^1 =0$, so that propagation is entirely in the
$(x^2, x^3$)-plane and the set-up is planar.

The corresponding wave amplitude $\phi$ is therefore
\begin{eqnarray}
 \phi(x) &=& \frac{e^{-i k_\sigma x^\sigma}}{r}\left\{
 1-\frac{1}{2}\left[\int_S^x dz^0 \gamma_{00,2} (x^0-z^0) \Pi^2 +
 \int_S^x
 dz^0 \gamma_{00, 3}(x^0-z^0) \Pi^3-\int_S^x dz^0 \gamma_{00,2}
 (x^2-z^2)\Pi^0
 \right.\right. \nonumber \\
 & - &\int_S^x dz^0 \gamma_{00, 3}(x^3-z^3) \Pi^0 + \int_S^x dz^2 \gamma_{22,3}(x^2-z^2)\Pi^3+
 \int_S^x dz^3 \gamma_{33,2}(x^3-z^3)
 \Pi^2-\int_S^x dz^2 \gamma_{22,3}
 (x^3-z^3)\Pi^2  \nonumber\\
  & - & \left. \left. \int_S^x dz^3 \gamma_{33,2}(x^2-z^2) \Pi^3\right] +
  \frac{1}{2}\left[\int_S^x dz^0 \gamma_{00}
 \Pi^0+\int_S^x dz^2 \gamma_{22} \Pi^2+\int_S^x dz^{3}\gamma_{33}\Pi^{3}\right] \right\}\,, \label{*}
\end{eqnarray}
where $\Pi^0=-i k$, $\Pi^i = -ik^i $, and we have taken into account
the fact that $\gamma_{11}$ plays no role in the planar arrangement
chosen. The phase must now be calculated along the different paths
SP+PO and SL+LO taking into account the values of $\Pi^i$ in the
various intervals.
\begin{figure}
\centering
\includegraphics[width=0.9\textwidth]{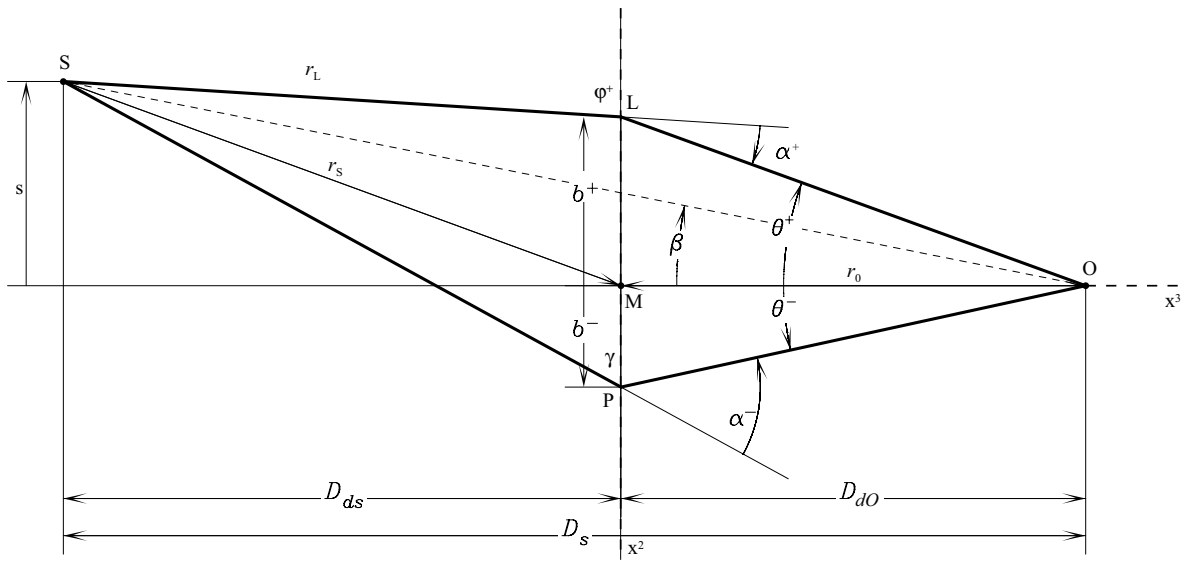}
\caption{\label{fig:Lensing} Geometry of a two-image gravitational
lens. The solid lines represent the particle paths between the
particle source at $S$ and the observer at $O$. $M$ is the
spherically symmetric gravitational lens. $S, M, O$ and the particle
paths lie in the same plane.}
\end{figure}

The total change in phase is \cite{pap3}
 \[
 \Delta \tilde{\phi}= \Delta \tilde{\phi}_{SL} +\Delta \tilde{\phi}_{LO}-\Delta \tilde{\phi}_{SP}-
 \Delta \tilde{\phi}_{PO}\,.
 \]
All integrations in (\ref{*}) can be performed exactly and the
results can be expressed in terms of physical variables
 $r_s$, $r_0$, $b^+$, $b^-$, and $s$ or lensing variables $D_s$,
 $D_{ds}$, $D_d$, $\theta^+$, $\theta^-$, and $\beta$. We find
\begin{eqnarray}\label{FR}
\Delta\tilde{\phi} & = & \tilde{y} \left\{\ln \left(-\sqrt{D_{dS}^2
+\left(s + b^-\right)^2} + b^- \cos\gamma + r_S \right)- \ln
\left(b^- \left(1+ \cos\gamma\right)\right)
\right.\nonumber \\
& + & \ln \left(b^+ \left(1-\cos\varphi^+ \right)\right)-
\ln\left(r_S -r_L-b^+\cos\varphi^+\right)
\nonumber \\
& + & \ln \left( b^- + r_0 \cos\theta^- -\sqrt{b^{-\,2}+
r_0^2}\right)-\ln \left(r_0\left(1+ \cos\theta^-\right)\right)
\nonumber \\
& + & \left.\ln\left(r_0\left(1+ \cos\theta^+\right)\right)-
\ln\left(b^+ +r_0 \cos\theta^+ -\sqrt{b^{+\,2}+r_0^2}\right)
\right\} \,,
\end{eqnarray}
where $ r_S^2 = b^{+\,2}+r_L^2+2b^+r_L\cos\varphi^+$\,,
$r_L^2=D_{dS}^2 +(s-b^+)^2$\,, $ \varphi^+ +\alpha^+ +\alpha^-
+\gamma -\theta^+ -\theta^- =\pi$ and $ \tilde{y}= 2GMk$.

As an example, let us consider the simpler case $ b^{+}= b^{-}\equiv
b\,, \theta^{+}=\theta^{-}\equiv\theta\,, s = 0\,, r_{L}=\sqrt{b^2
+r_0^2}$, from which we obtain (see Fig.1) $r_{S}= r_0\,, \cos(\pi-
\varphi^{+})= \cos\gamma=b/\sqrt{r_{0}^{2}+b^{2}}\, \tan \theta =
b/r_0$.

A simple calculation shows that the probability density of finding a
photon, or graviton, at $ O$ is typically
\begin{equation}\label{PD}
\phi\phi^* \propto \cos^2 \frac{\Delta\tilde{\phi}}{2}\,.
\end{equation}

\section{Muon $ g-2 $ experiments}
\setcounter{equation}{0}

Measurements of the interaction of spin with rotation have been
carried out in the case of photons using signals from global
positioning system satellites \cite{ashby} and data published in
\cite{versuve} can be re-interpreted \cite{mashhoon3} as due to the
coupling of Earth's rotation to the nuclear spins in mercury. The
spin-rotation effect is also consistent with a small depolarization
of electrons in storage rings \cite{bell}. We show below that the
same coupling is of particular interest in experiments with storage
rings. It is essential in getting the correct $ g$-dependence in $
g-2$ experiments \cite{paplamb} and eliminates the need of \emph{ad
hoc}, phenomenological arguments.

The effect is conceptually important. It extends to the quantum
level the classical coupling of rotation to the intrinsic angular
momentum of a body and simplifies the treatment of rotational
inertia.

It also yields different potentials for different particles and for
different spin states and can not, therefore, be considered
universal \cite{mashhoon2}.

Before discussing its connection with $ g-2$ experiments, it is
useful to briefly recall the usual experimental setup.

The experiment \cite{bailey,farley} involves muons in a storage ring
a few meters in diameter, in a uniform vertical magnetic field.
Muons on equilibrium orbits within a small fraction of the maximum
momentum are almost completely polarized with spin vectors pointing
in the direction of motion. As the muons decay, those electrons
projected forward in the muon rest frame are detected around the
ring. Their angular distribution thence reflects the precession of
the muon spin along the cyclotron orbits.

The calculations are performed in the rotating frame of the muon and
do not therefore require a relativistic treatment of inertial spin
effects \cite{ryder}  . Then the vierbein formalism yields
$\Gamma_i=0$ and
\begin{equation}\label{1}
  \Gamma_0=-\frac{i}{2}\,
  a_i\sigma^{0i}-\frac{1}{2}\,\omega_i\sigma^i\,,
\end{equation}
where $a_i$ and $\omega_i$ are the three-acceleration and
three-rotation of the observer and, in the chiral representation of
the usual Dirac matrices,
 \[
 \sigma^{0i}\equiv\frac{i}{2}\, [\gamma^0, \gamma^i]=i
 \left(\matrix{ \sigma^i & 0 \cr
                0 & -\sigma^i \cr }\right)\,.
 \]
The second term in (\ref{1}) represents the Mashhoon term. The first
term drops out. In fact, by symmetrization we obtain
\begin{equation}\label{simp}
\left(\vec{a}\cdot \vec{x}\right)\left(\vec{\alpha}\cdot
\vec{p}\right)= \frac{1}{2}\left[\left(\vec{a}\cdot
\vec{x}\right)\left(\vec{\alpha}\cdot
\vec{p}\right)+\left(\vec{\alpha}\cdot
\vec{p}\right)\left(\vec{a}\cdot
\vec{x}\right)\right]+\frac{i\hbar}{2}\left(\vec{a}\cdot
\vec{\alpha}\right).
\end{equation}
 The last term in (\ref{simp}) and the first term in
(\ref{1}) therefore cancel each other. The remaining first order
contributions in $a_i$ and $\omega_i$ to the Dirac Hamiltonian add
up to \cite{hehlni,singh}
\begin{eqnarray}\label{hamiltonian-Ni}
  H &\approx & {\vec \alpha}\cdot {\vec p}+m\beta+\frac{1}{2}
  [({\vec a}\cdot {\vec x})({\vec p}\cdot {\vec \alpha})+
  ({\vec p}\cdot {\vec \alpha})({\vec a}\cdot {\vec x})] \\
  & & -{\vec \omega}\cdot \left({\vec L}+\frac{{\vec \sigma}}{2}\right)\,.
  \nonumber
\end{eqnarray}
For simplicity all quantities in $H$ are taken to be
time-independent. They are referred to a left-handed tern of axes
rotating  about the $x_2$-axis in the clockwise direction of motion
of the muons. The $x_3$-axis is tangent to the orbits and in the
direction of the muon momentum. The magnetic field is $B_2=-B$. Only
the Mashhoon term and the magnetic moment interaction then couple
the helicity states of the muon. The remaining terms contribute to
the overall energy $E$ of the states, and we indicate by $H_0$ the
corresponding part of the Hamiltonian.

Before decay the muon states can be represented as
\begin{equation}\label{2}
  |\psi(t)>=a(t)|\psi_+>+b(t)|\psi_->\,,
\end{equation}
where $|\psi_+>$ and $|\psi_->$ are the right and left helicity
states of the Hamiltonian $H_0$ and satisfy the equation
 \[
 H_0|\psi_{+,-}>=E|\psi_{+,-}>\,.
 \]
The total Hamiltonian reduces effectively to $H_{eff}=H_0+H'$, where
\begin{equation}\label{3}
  H'=-\frac{1}{2}\,\omega_2\sigma^2+\mu B\sigma^2\,.
\end{equation}
$\displaystyle{\mu=\left(1+\frac{g-2}{2}\right)\mu_0}$ represents
the total magnetic moment of the muon and $\mu_0$ is the Bohr
magneton. Electric fields used to stabilize the orbits and stray
radial electric fields can also affect the muon spin. Their
effects can however be cancelled by choosing an appropriate muon
momentum \cite{farley} and will not be considered in what follows.

The coefficients $a(t)$ and $b(t)$ in (\ref{2}) evolve in time
according to
\begin{equation}\label{4}
  i\frac{\partial}{\partial t} \left(\matrix{ a(t) \cr
                b(t) \cr }\right)=M \left(\matrix{ a(t) \cr
                b(t) \cr }\right)\,,
\end{equation}
where $M$ is the matrix
\begin{equation}\label{5}
  M= \left[\matrix{ E-i\displaystyle{\frac{\Gamma}{2}} &
            \displaystyle{i\left(\frac{\omega_2}{2}-\mu B\right)}\cr
                \displaystyle{-i\left(\frac{\omega_2}{2}-\mu B\right)} &
                E-i\displaystyle{\frac{\Gamma}{2}} \cr }\right]
\end{equation}
and $\Gamma$ represents the width of the muon. The non-diagonal form
of $M$ (when $B=0$) implies that rotation does not couple
universally to matter.

 $M$ has eigenvalues
\begin{eqnarray}
 h_1 &=& E-i\frac{\Gamma}{2}+\frac{\omega_2}{2}-\mu B \,, \nonumber \\
 h_2 &=& E-i\frac{\Gamma}{2}-\frac{\omega_2}{2}+\mu B \,, \nonumber
\end{eqnarray}
and eigenstates
\begin{eqnarray}
 |\psi_1> &=&
 \frac{1}{\sqrt{2}}\,\left[i|\psi_+>+|\psi_->\right]\,, \nonumber
 \\
 |\psi_2> &=& \frac{1}{\sqrt{2}}\,\left[-i|\psi_+>+|\psi_->\right]
 \,. \nonumber
\end{eqnarray}
The muon states that satisfy (\ref{4}), and the condition $|\psi
(0)>=|\psi_->$ at $t=0$, are
\begin{eqnarray} \label{6a}
 |\psi(t)> &=& \frac{e^{-\Gamma t/2}}{2} e^{-iEt}
 \left\{
 i\left[e^{-i{\tilde \omega} t}
 -e^{i{\tilde \omega} t}\right]|\psi_+> \right.
 \\
 & & \left. + \left[e^{-i{\tilde \omega} t}
 +e^{i{\tilde \omega} t}\right]|\psi_-> \right\}
 \,, \nonumber
\end{eqnarray}
where
 \[
 {\tilde \omega}\equiv \frac{\omega_2}{2}-\mu B\,.
 \]
The spin-flip probability therefore is
\begin{eqnarray}\label{7}
  P_{\psi_-\to \psi_+}&=&|<\psi_+|\psi(t)>|^2 \\
     &=& \frac{e^{-\Gamma
  t}}{2}[1-\cos(2\mu B-\omega_2) t]\,. \nonumber
\end{eqnarray}
The $\Gamma$-term in (\ref{7}) accounts for the observed exponential
decrease in electron counts due to the loss of muons by radioactive
decay \cite{farley}.

The spin-rotation contribution to $P_{\psi_-\to \psi_+}$ is
represented by $\omega_2$ which is the cyclotron angular velocity
$\displaystyle{\frac{eB}{m}}$ \cite{farley}. The spin-flip angular
frequency is then
 \begin{eqnarray}\label{omegafin}
 \Omega&=&2\mu B-\omega_2 \\
 &=&\left(1+\frac{g-2}{2}\right)\frac{eB}{m}-
 \frac{eB}{m} \nonumber \\
 &=& \frac{g-2}{2}\frac{eB}{m}\,, \nonumber
 \end{eqnarray}
which is precisely the observed modulation frequency of the
electron counts \cite{picasso}. This result is independent of the
value of the anomalous magnetic moment of the particle. The
cancellation of the Dirac value of the magnetic moment
contribution by the Mashhoon term must therefore take place for
all spin-1/2 particles in a similar physical set-up.  Hence, it is
the spin-rotation coupling that generates the correct $g-2$ factor
in $\Omega$ by exactly cancelling, in $2\mu B$, the much larger
contribution $\mu_0$ that fermions with no anomalous magnetic
moment produce. The cancellation is made possible by the
non-diagonal form of $M$ and is therefore a direct consequence of
the violation of the equivalence principle. It has, of course,
been argued that this principle does not hold true in the quantum
world \cite{lammer}.

\section{Constraints on the C and P symmetries}
\setcounter{equation}{0}

 Recently, discrepancies between the
experimental and standard model values of $ a_{\mu}$ have been
observed with very high accuracy \cite{brown2}. The most precise
data yet give $b =a_{\mu}(exp)-a_{\mu}(SM)=26\times 10^{-10}$ for
the negative muon \cite{brown3}, and $d
=a_{\mu}(exp)-a_{\mu}(SM)=33\times 10^{-10}$ for the positive muon
\cite{brown4}. This discrepancy can be used to set upper limits on
$P$ and $T$ invariance violations in spin-rotation coupling
\cite{lambpap,lambpap1}.

The possibility that discrete symmetries in gravitation be not
conserved has been discussed in the literature. Attention has in
general focused on the potential
\begin{equation}\label{1a}
U(\vec{r})=\frac{GM}{r}\left[\alpha_{1}\vec{\sigma}\cdot
\hat{r}+\alpha_{2}\vec{\sigma}\cdot \vec{v}+\alpha_{3}\hat{r}\cdot(
\vec{v}\times \vec{\sigma})\right],
\end{equation}
which applies to a particle of generic spin $\vec{\sigma}$. The
first term, introduced by Leitner and Okubo \cite{leitner}, violates
the conservation of $P$ and $T$. The same authors determined the
upper limit $\alpha_{1}\leq 10^{-11}$ from the hyperfine splitting
of the ground state of hydrogen. The upper limit $\alpha_{2}\leq
10^{-3}$ was determined in \cite{almeida} from SN 1987A data. The
corresponding potential violates the conservation of $P$ and $C$.
Conservation of $C$ and $T$ is violated by the last term, while
(\ref{1a}), as a whole, conserves $CPT$. There is, as yet, no upper
limit on $\alpha_{3}$. These studies are extended here to the
Mashhoon term.

Before decay, the muon states can be represented as in (\ref{2})
where $\mid\psi_{+}>$ and $\mid\psi_{-}>$ again are the right and
left helicity states of the Hamiltonian $H_{0}$ defined in the
previous section.

Assume now that the coupling of rotation to $\mid\psi_{+}>$ differs
in strength from that to $\mid\psi_{-}>$. Then the Mashhoon term can
be modified by means of a matrix $A=\left(\matrix{\kappa_{+}&0\cr
0&\kappa_{2-}\cr}\right)$ that reflects the different coupling of
rotation to the two helicity states. The total Hamiltonian now is
$H_{eff}=H_{0}+H'$, where
\begin{equation}\label{4a}
H'=-\frac{1}{2}A \omega_{2}\sigma_{2}+\mu B\sigma_{2}.
\end{equation}
A violation of $P$ and $T$ in (\ref{4a}) would arise through
$\kappa_{+}-\kappa_{-}\neq 0$. The constants $\kappa_{+}$ and
$\kappa_{-}$ are assumed to differ from unity by small amounts
$\epsilon_{+}$ and $\epsilon_{-}$.

The coefficients $a(t)$ and $b(t)$ in (\ref{2}) evolve in time
according to
\begin{equation}\label{5a}
i\frac{\partial}{\partial t}\left(\matrix{a(t)\cr
b(t)\cr}\right)=M\left(\matrix{a(t)\cr b(t)\cr}\right),
\end{equation}
where
\begin{equation}\label{6b}
M=\left(\matrix{E-i\frac{\Gamma}{2}&
i\left(\kappa_{+}\frac{\omega_{2}}{2}-\mu B\right)\cr
-i\left(\kappa_{-}\frac{\omega_{2}}{2}-\mu B\right)&
E-i\frac{\Gamma}{2}\cr}\right),
\end{equation}
and $\Gamma$ represents, as before, the width of the muon. The
spin-rotation term is off-diagonal in (\ref{6b}) and does not
therefore couple to matter universally. It violates Hermiticity as
shown in \cite{papini2} and, in a general way, by Scolarici and
Solombrino \cite{scolarici}. It also violates $T$, $P$ and $PT$,
while nothing can be said about $CPT$ conservation which requires
$H_{eff}$ to be Hermitian. Because of the non-Hermitian nature of
(\ref{4a}), one expects $\Gamma$ itself to be non-Hermitian. The
resulting corrections to the width of the muon are, however, of
second order in the $\epsilon$'s and are neglected.

$M$ has eigenvalues
\begin{eqnarray}
h_{1}&=&E-i\frac{\Gamma}{2}+R \nonumber \\
h_{2}&=&E-i\frac{\Gamma}{2}-R,
\end{eqnarray}
where
\begin{equation}
R=\sqrt{\left(\kappa_{+}\frac{\omega_{2}}{2}-\mu
B\right)\left(\kappa_{-}\frac{\omega_{2}}{2}-\mu B\right)},
\end{equation}
and eigenstates
\begin{eqnarray}
|\psi_{1}>&=&b_{1}\left[\eta_{1}|\psi_{+}>+|\psi_{-}>\right],\nonumber \\
|\psi_{2}>&=&b_{2}\left[\eta_{2}|\psi_{+}>+|\psi_{-}>\right].
\end{eqnarray}
One also finds
\begin{eqnarray}
|b_{1}|^{2}&=&\frac{1}{1+|\eta_{1}|^{2}}\nonumber \\
|b_{2}|^{2}&=&\frac{1}{1+|\eta_{2}|^{2}}
\end{eqnarray}
and
\begin{equation}
\eta_{1}=-\eta_{2}=\frac{i}{R}\left(\kappa_{+}\frac{\omega_{2}}{2}-\mu
B\right).
\end{equation}
Then the muon states (\ref{2}) are
\begin{eqnarray}
|\psi(t)>&=&\frac{1}{2}e^{-iEt-\frac{\Gamma t}{2}}[-2i\eta_{1} \sin
Rt |\psi_{+}>+
\nonumber \\
         & &2\cos Rt
         |\psi_{-}>],
\end{eqnarray}
where the condition $|\psi(0)>=|\psi_{-}>$ has been applied. The
spin-flip probability is therefore
\begin{eqnarray}\label{9}
P_{\psi_{-}\rightarrow \psi_{+}}&=&|<\psi_{+}|\psi(t)>|^{2}\nonumber
\\&=&\frac{e^{-\Gamma t}}{2\left(1 + \kappa_{-}\omega_{2}-2\mu B\right)}
\left[1-\cos \left(2Rt\right)\right].
\end{eqnarray}
When $ \kappa_{+}=\kappa_{-}=1$, (\ref{9}) yields \cite{lambpap}
\begin{equation}\label{10}
P_{\psi_{-}\rightarrow \psi_{+}}=\frac{e^{-\Gamma
t}}{2}\left[1-\cos\left(a_{\mu} \frac{eB}{m}t\right)\right],
\end{equation}
that provides the appropriate description of the spin-rotation
contribution to the spin-flip transition probability. Notice that
the case $\kappa_{+}=\kappa_{-}=0$ (no spin-rotation coupling)
yields
\begin{equation}\label{A1}
P_{\psi_{-}\rightarrow \psi_{+}}=\frac{e^{-\Gamma
t}}{2}\left[1-\cos(1+a_{\mu})\frac{eB}{m}\right]
\end{equation}
and does not therefore agree with the results of the $g-2$
experiments. Hence the necessity of accounting for spin-rotation
coupling whose contribution cancels the factor $\frac{eB}{m}$ in
(\ref{A1}).

Substituting $\kappa_{+}=1+\epsilon_{+},
 \kappa_{-}=1+\epsilon_{-}$ into (\ref{9}), one finds
\begin{equation}\label{11}
P_{\psi_{-}\rightarrow\psi_{+}}=\frac{e^{-\Gamma
t}}{2}\frac{2\left(\epsilon_{+}-a_{\mu}\right)}{\epsilon_{+}+\epsilon_{-}-2a_{\mu}}
\left[1-\cos\left(t\frac{eB}{m}\sqrt{\left(\epsilon_{+}-
a_{\mu}\right)\left(\epsilon_{-}-a_{\mu}\right)}\right)\right].
\end{equation}

One may attribute the discrepancy between $a_{\mu}(exp)$ and
$a_{\mu}(SM)$ to a violation of the conservation of the discrete
symmetries by the spin-rotation coupling term in (\ref{4}). The
upper limit on the violation of $P,T$ and $PT$ is derived from
(\ref{11}) assuming that the deviation from the current value of
$a_{\mu}(SM)$ is wholly due to $\epsilon_{\pm}$. The upper limit is
therefore $26\times 10^{-10}$ in the case of negative muons
\cite{brown3} and of $ 33 \times 10^{-10} $ for positive muons
\cite{brown4}. At the same time the two values of
$a_{\mu}(exp)-a_{\mu}(SM)$ can be thought of as due to a different
coupling strength between rotational inertia and the two helicity
states of the muon. Then the values of $ \epsilon_{+}$ and $
\epsilon_{-}$ can be determined from $ cos\left(2 R t\right)$ in
(\ref{9}) according to the equations

\begin{equation}\label{11a}
\left(a_{\mu_{+}}-
\epsilon_{+}\right)\left(a_{\mu_{+}}-\epsilon_{-}\right)=b^{2}
\end{equation}
and
\begin{equation}\label{11b}
\left(a_{\mu_{-}}-
\epsilon_{+}\right)\left(a_{\mu_{-}}-\epsilon_{-}\right)=d^{2}\,.
\end{equation}
Equations (\ref{11a}) and (\ref{11b}) have the approximate solutions
\begin{equation}\label{11c}
\epsilon_{+}\simeq \frac{a_{\mu_{+}}+
a_{\mu_{-}}}{2}-\frac{d^{2}-b^{2}}{2\left(a_{\mu_{-}}-a_{\mu_{+}}\right)}
\end{equation}
and
\begin{equation}\label{11d}
\epsilon_{-}\simeq a_{\mu_{+}}+ \frac{2
b^{2}\left(a_{\mu_{-}}-a_{\mu_{+}}\right)}{\left(a_{\mu_{-}}-a_{\mu_{+}}\right)^{2}+\left(d^{2}-b^{2}\right)}.
\end{equation}
More precise, numerical solutions give $ \epsilon_{+}\simeq 11659189
\cdot 10^{-10},  \epsilon_{-} \simeq 11659152 \cdot 10^{-10}$ and $
\Delta\epsilon \equiv \epsilon_{+}-\epsilon_{-}\simeq 37.65878 \cdot
10^{-10}$. These values are significant in view of the precision
with which $ a_{\mu\pm}, b, d$ have been determined. It then follows
that the coupling of rotation to positive helicity is larger than
that to negative helicity, which agrees with the value $ \xi < 1$
for both $a_{\mu}= a_{\mu+}$ and $ a_{\mu}=a_{\mu-}$. This also
means that the violation of $ P$ and $ T$ is relatively stronger for
positive helicity at a level $ \Delta\epsilon \simeq 3.7\cdot
10^{-9}$ and that the spin-rotation interaction is an inherent
source of $ P$ and $T$ violation.

\section{Neutrino Helicity Transitions}
\setcounter{equation}{0}

In this section, it is convenient to write the left and right
neutrino wave functions in the form
\begin{equation}\label{Psi0}
 \Psi_0 (x) = \nu_{0L,R}e^{-ik_\alpha x^\alpha}=\sqrt{\frac{E+m}{2E}}
  \left(\begin{array}{c}
                \nu_{L, R} \\
                 \frac{{\bbox \sigma}\cdot {\bf k}}{E+m}\, \nu_{L, R} \end{array}\right)
                 \,e^{-ik_\alpha x^\alpha}\,,
\end{equation}
where $\bbox \sigma=(\sigma^1, \sigma^2, \sigma^3)$ represents the
Pauli matrices, $\nu_{L,R}$ are eigenvectors of $\bbox
\sigma\cdot\bbox k$ corresponding to negative and positive helicity
and ${\bar \nu}_{0\, L, R}(k)\equiv \nu_{0\, L, R}^\dagger
(k)\gamma^{\hat 0}$, ${\nu}_{0\, L, R }^\dagger(k)\nu_{0\, L,
R}(k)=1$. This notation already takes into account the fact that if
$ \nu_{\pm}$ are the helicity states, then we have $ \nu_{L}\simeq
\nu_{-}, \, \nu_{R}\simeq \nu_{+}$ for relativistic neutrinos.

In general, the spin precesses during the motion of the neutrino.
This can be seen, for instance, from the contribution $ \Phi_{s}$ in
$ \Phi_{T}$. The expectation value of the contribution of $
\Gamma_{0}$ to the effective mechanical momentum can in fact be
re-written in the form
\begin{equation}\label{Omega1}
 \frac{1}{2}\Psi_{0}^{\dag}
\vec{\Omega}\cdot \vec{\sigma}\Psi_{0},
\end{equation}
where $ \vec{\Omega} \equiv
\frac{GMR^{2}}{5r^{3}}\left(1-\frac{3z^2}{r^2}\right)\vec{\omega}$.
Equation (\ref{Omega1}) represents the spin-rotation coupling for
the Lense-Thirring metric. Here rotation is provided by the
gravitational source, rather than by the particles themselves.

We now study the helicity flip of one flavor neutrinos as they
propagate in the gravitational field produced by a rotating mass
\cite{punzi}. The neutrino state vector can be written as
\begin{equation}\label{neutrinoRL}
 |\psi(\lambda) \rangle = \alpha(\lambda) |\nu_R\rangle
 +\beta(\lambda) |\nu_L\rangle\,,
\end{equation}
where $|\alpha|^2+|\beta|^2=1$ and $\lambda$ is an affine
parameter along the world-line. In order to determine $\alpha $
and $\beta $, we can write (\ref{PsiSolution}) as

\begin{equation}\label{PsiSolution3}
  |\psi(\lambda)\rangle = {\hat T}(\lambda) |\psi_0(\lambda)\rangle\,,
\end{equation}
where
\begin{equation}\label{Texpression}
  {\hat T}=-\frac{1}{2m}\left(-i\gamma^\mu(x){\cal
  D}_\mu-m\right)e^{-i\Phi_T}\,,
\end{equation}
and $|\psi_0(\lambda)\rangle $ is the corresponding solution in
Minkowski space-time. The latter can be written as
\begin{equation}\label{psizero}
|\psi_{0}(\lambda)\rangle = e^{-ik\cdot x}
\left[\alpha(0)|\nu_{R}\rangle + \beta(0)|\nu_{L}\rangle \right]\,.
\end{equation}
Strictly speaking, $ |\psi(\lambda)\rangle $ should also be
normalized. However, it can be shown \cite{punzi} that $
\alpha(\lambda) $ is already of $ \mathcal{O}(\gamma_{\mu\nu})$, can
only produce higher order terms and is therefore unnecessary in this
calculation. From (\ref{neutrinoRL}), (\ref{PsiSolution3}) and
(\ref{psizero}) we obtain
\begin{equation}\label{alpha}
\alpha(\lambda)=\langle
\nu_{R}|\psi(\lambda)\rangle=\alpha(0)\langle\nu_{R}|\hat{T}|\nu_{R}\rangle
+ \beta(0)\langle \nu_{R}|\hat{T}|\nu_{L}\rangle\,.
\end{equation}
An equation for $ \beta $ can be derived in an entirely similar way.

If we consider neutrinos which are created in the left-handed state,
then $|\alpha(0)|^2=0,  |\beta(0)|^{2}=1 $, and we obtain
\begin{equation}\label{PLR}
  P_{L\rightarrow R}=|\alpha(\lambda)|^2=\left|\langle \nu_{R}|\hat{T}|\nu_{L}\rangle\right|^{2}
  =\left|\int_{\lambda_0}^\lambda\langle \nu_R| {\dot x}^\mu \partial_\mu {\hat
  T}|\nu_L\rangle d\lambda\right|^2\\,
\end{equation}
where $\dot{x}^{\mu}=k^{\mu}/m$. As remarked in \cite{cardall}, $
\dot{x}^{\mu}$ need not be a null vector if we assume that the
neutrino moves along an "average" trajectory. We also find, to
lowest order,
\begin{eqnarray}\label{deT}
  \partial_{\mu}\hat{T}&=&\frac{1}{2m}\left(-i2m\Phi_{G,\mu}-i(\gamma^{\hat{\alpha}}k_{\alpha}+m)\Phi_{s,\mu}
  +\gamma^{\hat{\alpha}}(h^{\beta}_{\hat{\alpha},\mu}k_{\beta}+\Phi_{G,\alpha\mu})\right)\\
  \Phi_{s,\lambda}&=&\Gamma_{\lambda}\,,\quad
  \Phi_{G,\alpha\mu}=k_{\beta}\Gamma^{\beta}_{\alpha\mu}\,,\quad
  \nu^{\dagger}_0(\gamma^{\hat{\alpha}}k_{\alpha}+m)=2E\nu^{\dagger}_{0}\gamma^{\hat{0}}\nonumber\,,
\end{eqnarray}
where $\Gamma^{\beta}_{\alpha\mu}$ are the usual Christoffel
symbols, and
\begin{equation}\label{RTL}
 \langle \nu_R |{\dot x}^\mu \partial_\mu {\hat T}|\nu_L \rangle
 =\frac{E}{m}\left[
 -i \, \frac{k^\lambda}{m} {\bar \nu}_R\Gamma_\lambda \nu_L+
 \frac{k^\lambda k_\mu}{2mE}\, (h^\mu_{{\hat \alpha},\,
 \lambda}+\Gamma^\mu_{\alpha\lambda})\nu_R^\dagger \gamma^{\hat
 \alpha}\nu_L\right].
\end{equation}
In what follows, we compute the probability amplitude (\ref{RTL})
for neutrinos propagating along the $z$ and the $x$ directions
explicitly.

\subsection{Propagation along $z$}

For propagation along the $z$-axis, we have $k^0=E$ and $k^3\equiv
k\simeq E(1-m^2/2E^2)$. As in Section III, we choose $y=0,\, x=b $.
We get
\begin{eqnarray}\label{firstcontrib}
  -i\frac{k^\lambda}{m}\bar{\nu}_R\Gamma_\lambda\nu_L&=&\frac{k}{m}\phi_{,1}+i\frac{m}{4E}h_{2,3}\,\,,\\
    \frac{k^\lambda k_\mu
    }{2mE}\, (h^\mu_{{\hat \alpha},\,
    \lambda}+\Gamma^\mu_{\alpha\lambda})\nu_R^\dagger \gamma^{\hat
    \alpha}\nu_L&=&-\frac{k}{2m}\left(1+\frac{k^2}{E^2}\right)\,.\nonumber
\end{eqnarray}
Summing up, and neglecting terms of $ {\cal O}(m/E)^2$,
(\ref{RTL}) becomes
\begin{equation}\label{RTL1}
\langle\nu_R|\dot{x}^{\mu}\partial_{\mu}\hat{T}|\nu_L\rangle=\frac{1}{2}\phi_{,1}+\frac{i}{4}h_{2,3}\,.
\end{equation}
The contributions to ${\cal O}((E/m)^2)$ vanish. As a consequence
\begin{equation}\label{dadz}
  \frac{d\alpha}{dz}\simeq\frac{m}{E}\frac{d\alpha}{d\lambda}=
  \frac{m}{E}\left(\frac{1}{2}\phi_{,1}+\frac{i}{4}h_{2,3}\right)\,,
\end{equation}
and the probability amplitude for the $ \nu_{L}\rightarrow\nu_{R}$
transition is of ${\cal O}(m/E)$, as expected.

Integrating (\ref{dadz}) from $-\infty$ to $z$, yields
\begin{eqnarray}\label{alfa-z}
  \alpha&\simeq&\frac{m}{E}\left[\frac{1}{2}\int_{-\infty}^zdz\phi_{,1}+\frac{i}{4}h_2(z)\right]\\
  &=&\frac{m}{E}\frac{GM}{2b}\left[1+\frac{z}{r}-i\frac{2\omega
  R^2b^2}{5r^3}\right]\,.\nonumber
\end{eqnarray}
It also follows that
\begin{equation}\label{|alfa-z|}
  P_{L\rightarrow R}(-\infty,z)\simeq \left(\frac{m}{E}\right)^2\left(\frac{GM}{2b}\right)^2
  \left[\left(1+\frac{z}{r}\right)^2+\left(\frac{2\omega b^2R^2}{5r^3}\right)^2\right]\,.
\end{equation}
The first of the two terms in (\ref{|alfa-z|}) comes from the mass
of the gravitational source. The second from the source's angular
momentum and vanishes for $r\rightarrow\infty$ because  the
contribution from $-\infty $ to 0 exactly cancels that from 0 to $
+\infty $. In fact, if we consider neutrinos propagating from 0 to $
+\infty $, we obtain
\begin{equation}\label{alfa infinito}
  P_{L\rightarrow R}(0,+\infty)\simeq \left(\frac{m}{E}\right)^2\left(\frac{GM}{2b}\right)^2
  \left[1+\left(\frac{2\omega R^2}{5b}\right)^2\right]\,.
\end{equation}
According to semiclassical spin precession equations
\cite{montague}, there should be no spin motion when spin and $
\vec{\omega}$ are parallel as in the present case. This is a hint
that rotation of the source, rather than of the particles, should
produce a similar effect. The probabilities (\ref{|alfa-z|}) and
(\ref{alfa infinito}) mark therefore a departure from expected
results. They are however small of second order. Both expressions
vanish for $ m \rightarrow 0$, as it should for a stationary metric.
In this case, in fact, helicity is conserved \cite{mobed}. It is
interesting to observe that spin precession also occurs when $
\omega$ vanishes \cite{aldov,casini}. In the case of
(\ref{|alfa-z|}) the mass contribution is larger when $ b<
(r/R)\sqrt{\frac{5r}{2\omega}}$, which, close to the source, with
$b\sim r\sim R$, becomes $ R\omega < 5/2$ and is always satisfied.
In the case described by (\ref{alfa infinito}), the rotational
contribution is larger if $ b/R < 2\omega R/5$ which effectively
restricts the region of dominance to a narrow strip about the $
z$-axis in the equatorial plane, if the source is compact and $
\omega$ is relatively large.

\subsection{Propagation along $x$}

In this case, we take $k^0=E$, $k^1\equiv k\simeq E(1-m^2/2E^2)$. As
in Section III, the calculation can be simplified by assuming that
the motion is in the equatorial plane with $z=0$, $y=b$. We then
have
\begin{eqnarray}\label{contribx}
-i\frac{k^\lambda}{m}\bar{\nu}_R\Gamma_\lambda\nu_L&=&i\frac{k}{m}\phi_{,2}+i\frac{E^2+k^2}{4mE}h_{1,2}-i\frac{E^2-k^2}{4mE}h_{2,1}\,\,,\\
    \frac{k^\lambda k_\mu
    }{2mE}\, (h^\mu_{{\hat \alpha},\,
    \lambda}+\Gamma^\mu_{\alpha\lambda})\nu_R^\dagger \gamma^{\hat
    \alpha}\nu_L&=&-i\frac{k}{2m}\left(1+\frac{k^2}{E^2}\right)\phi_{,2}-i\frac{k^2}{2mE}h_{1,2}\,.\nonumber
\end{eqnarray}
Summing up, and neglecting terms of $ {\cal O}(m/E)^2$,
(\ref{RTL}) becomes
\begin{equation}\label{RTL1x}
\langle\nu_R|\dot{x}^{\mu}\partial_{\mu}\hat{T}|\nu_L\rangle=\frac{i}{2}\phi_{,2}+\frac{i}{4}(h_{1,2}-h_{2,1})\,.
\end{equation}
The contributions to ${\cal O}((E/m)^2)$ again vanish and we get
\begin{equation}\label{dadx}
\frac{d\alpha}{dx}\simeq\frac{m}{E}\frac{d\alpha}{d\lambda}=\frac{m}{E}\left[\frac{i}{2}
\phi_{,2}+\frac{i}{4}(h_{1,2}-h_{2,1})\right]\sim{\cal O}(m/E)\,.
\end{equation}
Integrating (\ref{dadx}) from $-\infty$ to $x$, we obtain
\begin{equation}\label{alfax}
  \alpha\simeq i\frac{m}{E}\frac{GM}{2b}\left(1-\frac{2\omega
  R^2}{5b}\right)\left(1+\frac{x}{r}\right)\,
\end{equation}
and
\begin{equation}\label{alfa2x}
  P_{L\rightarrow
  R}(-\infty,x)\simeq\left(\frac{m}{E}\right)^2\left(\frac{GM}{2b}\right)^2\left(1-\frac{2\omega
  R^2}{5b}\right)^2\left(1+\frac{x}{r}\right)^2\,.
\end{equation}
Obviously, the mass contribution is the same as for propagation
along the $z$-axis. However, the two cases differ substantially in
the behavior of the angular momentum term. In this case, in fact,
this term is even, so it does not vanish for $r\rightarrow\infty$.
If we consider neutrinos generated at $x=0$ and propagating to
$x=+\infty$, we find
\begin{equation}\label{pinfinitox}
P_{L\rightarrow
R}(0,+\infty)\simeq\left(\frac{m}{E}\right)^2\left(\frac{GM}{2b}\right)^2\left(1-\frac{2\omega
  R^2}{5b}\right)^2\,.
\end{equation}
The mass term is larger when $ \frac{2\omega R^{2}}{5b}< 1$. At the
poles $ b\sim R$ and the mass term dominates because the condition $
\omega R <5/2$ is always satisfied. The angular momentum
contribution prevails in proximity of the equatorial plane. The
transition probability vanishes at $ b=2\omega R^{2}/5$.

\section{\label{sec:5}Conclusions}

Covariant wave equations for massless and massive particles can be
solved exactly to first order in $ \gamma_{\mu\nu}$. The solutions
are covariant and invariant with respect to the gauge
transformations of the electromagnetic field and of $
\gamma_{\mu\nu}$ and are known when a solution of the free wave
equation is known. The external gravitational field only appears in
the phase of the wave function.

We have shown that the coupling of spin to inertia and gravitation
follow from the solutions given. This allows a unified treatment of
the interaction of gravity with spin and angular momentum without
requiring ad hoc procedures.

According to equations (\ref{l}) and (\ref{sol'}), the spin term $
S_{\alpha\beta}$ finds its origin in the skew-symmetric part of the
space-time connection. In the case of fermions, $ S_{\alpha\beta}$
is accounted for by the spinorial connection. The terms that contain
$ S_{\alpha\beta}$ gives rise to the Skrotskii effect for both
electromagnetic \cite{skr} and gravitational waves \cite{ramos}.

From the phases we have derived the geometrical optics of the
particles and verified that their deflection is that predicted by
general relativity. In addition, the background gravitational field
acts as a medium whose index of refraction can be calculated for any
metric from (\ref{gph}), (\ref{photonmomentum}) and $ n =
\tilde{k}/\tilde{k}_0$.

Because spin does not enter the examples given, the same results can
be equally applied to the gravitational lensing of gravitational
waves\cite{tak}.

A more detailed treatment of the geometrical optics of single flavor
neutrinos can be found in \cite{punzi} where we also calculate
corrections due to the neutrino mass. For propagation parallel to
the axis of rotation of the source, the rotation corrections vanish
at infinity. Not so for propagation perpendicular to the axis of
rotation.

We can finally conclude that the validity of covariant wave
equations in an inertial-gravitational context finds support in
experimental verifications of some of the effects they predict
\cite{COW,page,bonse}, in tests of the general relativistic
deflection of light rays and also in the phase wrap-up in global
position system measurements \cite{ashby}.

We have then asked ourselves the question whether, beside the phase
wrap-up in GPS, there is a wider role for spin-gravity coupling in
physics. In particular, we have considered muon $g-2$ experiments
and helicity transitions in neutrino physics. We have found that
spin-rotation coupling is largely responsible for producing the
correct $g-2$ factor in the spin-flip angular frequency $ \Omega$.
Measurements of this factor already provide the most stringent test
yet of Einstein's time dilation formula \cite{combley,cooper}.
However, muons in storage rings are also rotating quantum gyroscopes
and inertia must be an essential ingredient of their description in
experiments of high and ever increasing sensitivity. Possibly these
experiments also concern problems like the violation of the
equivalence principle in quantum mechanics and the conservation of
discrete symmetries in inertia-gravitation. We have, in fact, shown
that a slightly anomalous inertial contribution to (\ref{11}) at a
level of the values $ b =26\times 10^{-10}$ and $ d=33\times
10^{-10}$ of Section VII can produce violations of the discrete
symmetries at the same level. The upper limits on the violations of
$ P$ and $ T$ that can be reached by $ g-2$ experiments are in fact
as sensitive as those obtained by other means
\cite{schiff,leitner,dass,almeida}, but with an important
difference. The $g-2$ measurements are performed in strictly
controlled laboratory conditions rather than in astrophysical
situations.

In derivations based on the covariant Dirac equation, the coupling
of inertia and gravitation to spin is identical to that for orbital
angular momentum. A suggestive interpretation of this result is that
the internal distributions of the gravitational mass, associated
with the interaction, and of inertial mass, associated with the
angular momentum, equal each other. This is no longer so when $
\epsilon_{\pm}\neq 0$. There is almost a similarity, here, with the
electromagnetic case where $ g=2 $ is required by the Dirac
equation, but not by quantum electrodynamics. The deviations of $
\kappa_{+} $ and $ \kappa_{-}$ from unity that are consistent with $
g-2$ experiments are both of the order of $ a_{\mu}$, or $ \simeq
10^{-3}$, and differ from each other by $ \Delta\epsilon \simeq
3.7\cdot 10^{-9}$. While small values of $ \epsilon_{\pm} $ do not
give rise to measurable mass differences in macroscopic objects
\cite{papini2}, violations of the discrete symmetries can have
interesting astrophysical and cosmological implications.

Next, we have calculated the helicity transition amplitudes of
ultra-relativistic, single flavor neutrinos as they propagate in a
Lense-Thirring field. These transitions are interesting because at
high energies chirality states are predominantly helicity states
and right-handed neutrinos do not interact \cite{cai2,papini94}.
The transition probabilities are of $ {\cal O}(
\gamma_{\mu\nu}^{2})$. Two directions of propagation have again
been selected and the results contain contributions from both mass
and angular momentum of the source. The transitions also occur in
the absence of rotation or with spin parallel to rotation, which
is unexpected on semiclassical grounds. The mass contributions
predominate when the neutrinos propagate from $ r=0$ to $ r =
\infty $ (and matter effects are neglected), provided the impact
parameter $ b
>2 \omega R^{2}/5$. There is, however, a narrow region about the
axis of propagation in the equatorial plane where the $ \omega$
contribution is larger. The rotational contribution behaves
differently in the two cases. It vanishes as $ z\rightarrow \mp
\infty$ for propagation along $z$, but not so as $ x\rightarrow
\infty$ in the second case. In addition, when the neutrinos
propagate from $ x=0$ to $ x=\infty$, the mass term dominates in the
neighborhood of the poles, while the contribution of $ \omega$ is
larger close to the equator, with no attenuation at $ b=2\omega
R^{2}/5$.

In \cite{punzi} we have also calculated gravity induced,
two-flavor oscillations and derived the relative equation and
effective Hamiltonian. The transition probabilities do indeed
oscillate for the Lense-Thirring metric, and the curvature of
space-time enters the oscillation probability through the
gravitational red-shift of the local energy $ E_{l}$ and the
proper distance $ dl$.

The results presented in this paper can be applied to a number of
problems in astroparticle physics and cosmology \cite{dolgov}. For
instance, an interesting question is whether gravity induced
helicity and flavor transitions could effect changes in the ratio $
\nu_{e}:\nu_{\mu}:\nu_{\tau}$ of the expected fluxes at Earth.

Lepton asymmetry in the Universe \cite{dolgov} also is an
interesting problem. It is known that the active-sterile oscillation
of neutrinos can generate a discrepancy in the neutrino and
antineutrino number densities. The lepton number of a neutrino of
flavor $ f$ is defined by $
L_{f}=(n_{\nu_f}-n_{\bar{\nu}_f})/n_{\gamma}(T)$, where $ n_{\nu_f}
(n_{\bar{\nu}_f})$ is the number density of neutrinos
(antineutrinos) and $ n_{\gamma}(T)$ is the number density of
photons at temperature $ T$. As noted above, the gravitational field
generates transitions from left-handed (active) neutrinos to
right-handed (sterile) neutrinos. If, in primordial conditions,
(\ref{alfa infinito}) and (\ref{pinfinitox}) become larger, then
helicity transitions may contribute in some measure to lepton
asymmetry.

Finally, we have recently re-examined the behavior of the
spin-gravity interaction and found that gravity can distinguish
between chirality and helicity \cite{mobed}. We have also found
that the spin-gravity interaction can distinguish between Dirac
and Majorana wave packets \cite{singh2}. A spin-flip does in fact
change a Majorana neutrino into an antineutrino and behaves like a
charge conjugation operation.

A few words of caution must now be added.

The spin-gravity couplings discussed in our work make use of the
weak field approximation in which gravity enters as a
non-dynamical field. There are, of course, physical situations in
which this approach can be trusted and the general agreement
between the quasi-classical and the quantum mechanical approaches
has been established \cite{silenko}. One could then be tempted to
extend our approximation procedure to any order in the metric
deviation, as suggested by equations (\ref{n}), (\ref{phasephi})
and (\ref{phase}). This would however lead to inconsistencies that
can only be removed, as shown by Deser \cite{deser}, by making use
of the full non-linear apparatus of general relativity.

Furthermore, it is assumed, in calculating the phases induced by
gravity, that all possible particle paths reduce, in the average,
to the phase integration paths. This approximation worsens the
more "quantum mechanical" particles and gravity become.

An additional point concerns the use of the locality hypothesis in
replacing non-inertial frames with inertial ones. As shown by
Mashhoon \cite{mashh2,mashh3,MASHH}, this hypothesis has
limitations and important consequences for the measuring process.
For standard accelerated measuring devices, for instance, it
entails the introduction of a maximal acceleration
\cite{caian1,caian2,GP,scarp}. In the absence of inertial frames,
of a complete formulation of spin-gravity couplings in curved
space-time and of direct experimental observations, assigning a
role to curvature in spin related problems may then prove
difficult, though worthy of investigation.

A final question regards the validity of the equivalence principle
in spin-gravity interactions. In $g-2$ experiments the interaction
of spin with gravity depends on the relative direction of spin and
rotation of the source. As such it is not universal and may be
regarded as violating those formulations of the equivalence
principle that hinge on universality. But a more fundamental
violation has been introduced in Section VII where the strength of
the coupling itself depends on the helicity of the particle
\cite{lambpap,lambpapini}. This is a violation of the
post-Newtonian equivalence principle, recently discussed by
Silenko and Teryaev \cite{teryaev}, by which a particle's spin and
angular momenta precession frequencies coincide. This principle
must be the object of rigorous experimental verifications and the
authors themselves suggest a number of tests to find even more
precise upper limits than those determined by using muon $g-2$
experiments \cite{lambpap}. In a closely related paper
\cite{silenko}, the same authors show that the spin-gravity dipole
coupling term found by Obukhov \cite{obuk} using the
Eriksen-Korlsrud transformation \cite{EK} does not lead to
observable effects.

Recently, impressive technical developments in the field of masers
\cite{walsworth} have succeeded in placing an upper limit of
$10^{-27} GeV$ on violations of Lorentz and $CPT$ symmetries.
Higher sensitivities are expected to be reached in the near
future. Since the interaction energy for a spin-$1/2$ particle in
the rotation field of Earth is of the same order of magnitude,
these developments bode well for the physics of spin-gravity
interactions.

\end{document}